\documentclass[twocolumn,prl,amsmath,preprintnumbers,amssymb,showpacs]
              {revtex4}
\usepackage{graphicx,macros,psfrag}

\begin{document}

\preprint{LMU-ASC 30/09}
\title{Tension dynamics and viscoelasticity of extensible wormlike chains} \author{Benedikt Obermayer}
\author{Erwin Frey}\email{frey@physik.lmu.de} \affiliation{Arnold Sommerfeld Center and Center for NanoScience, Ludwig-Maximilians-Universit\"at M\"unchen, Theresienstr. 37, 80333 M\"unchen, Germany}
 \date{\today}

\begin{abstract}
  The dynamic response of prestressed semiflexible biopolymers is characterized by the propagation and relaxation of tension, which arises due to the near inextensibility of a stiff backbone. It is coupled to the dynamics of contour length stored in thermal undulations, but also to the local relaxation of elongational strain. We present a systematic theory of tension dynamics for stiff yet extensible wormlike chains. Our results show that even moderate prestress gives rise to distinct Rouse-like extensibility signatures in the high-frequency viscoelastic response.
\end{abstract}
\maketitle

Recent experiments have successfully linked the viscoelastic properties of living cells to the rheological behavior of prestressed biopolymer networks~\cite{fernandez-pullarkat-ott:06,gardel-etal:06,majumdar-etal:08,kasza-etal:09}. Single filaments within these networks are well described by the wormlike chain (WLC) model, where very stiff backbones are idealized as inextensible space curves~\cite{saito-takahashi-yunoki:67}, giving rise to a characteristic $f^{-1/2}$-divergence of the force $f$ required to attain full stretching~\cite{marko-siggia:95}. Here, ``stretching'' needs to be seen as a ``straightening'' of excess length stored in thermal contour undulations~\cite{seifert-wintz-nelson:96}, suggesting the phrase ``pulling out stored length''. If sudden forces are applied, stored length can be pulled out at first only from growing boundary layers of size $\lpar(t)$ near the ends due to longitudinal friction with the viscous solvent~\cite{seifert-wintz-nelson:96,ajdari-juelicher-maggs:97,everaers-juelicher-ajdari-maggs:99,brochard-buguin-degennes:99,hallatschek-frey-kroy:05}.  The precise time dependence of $\lpar(t)$ is influenced in a quite subtle way by the applied prestress~\cite{obermayer-etal:07}, and a non-homogeneous distribution of stored length along the contour correspondings to a non-uniform tension profile.

Clearly, as the prestress gets larger, an increasing contribution to the longitudinal extension stems from the microscopic elasticity of the backbone bonds and less from thermal undulations. Hence, for an extensible backbone the polymer's response is characterized by a local competition between destroying thermal stored length and creating elongational strain. While it has long been recognized that stretching modes of long and slender elastic rods relax extremely fast~\cite{soda:73}, in a non-equilibrium setup they do so only up to the scale where the much slower stored length dynamics has equilibrated. Especially for bead-spring simulations, where realistically stiff backbones often require unfeasibly short time steps, it is not immediately clear if and how  backbone stretching affects the longitudinal relaxation. In this report, we present a theory of tension dynamics for stiff yet finitely extensible wormlike chains. A brief comparison of extensible and inextensible polymer models is used to motivate the ensuing systematic derivation based on the inextensible analog presented in Ref.~\cite{hallatschek-frey-kroy:05}. We then calculate  viscoelastic response properties and show that even moderate prestress can give rise to distinct extensibility signatures reminiscent of a Rouse-like dynamics.

In the WLC model, the polymer backbone is idealized as continuous space curve $\bvec r(s)$. Contour undulations are penalized with a bending energy proportional to the squared local curvature. Strict inextensibility would require that $s$ be the arclength such that $\bvec r'(s)$ is a unit vector, and this hard constraint allows exact solutions only for special cases~\cite{saito-takahashi-yunoki:67}. In general, Lagrange multipliers of varying sophistication are used to enforce miscellaneous constraints of different rigidity~\cite{harris-hearst:66,winkler-reineker-harnau:94,ha-thirumalai:95,liverpool:05,munk-etal:06,hinczewski-etal:09}. Specifically, for a non-equilibrium scenario with non-uniform stored length dynamics, it is inevitable to use a local constraint~\cite{goldstein-langer:95,seifert-wintz-nelson:96,morse:98c,brochard-buguin-degennes:99,shankar-pasquali-morse:02,hallatschek-frey-kroy:05,obermayer-etal:09}, which is intuitively interpreted as backbone tension. In the case of an extensible backbone this tension arises naturally as a spring force~\cite{kierfeld-etal:04,netz:01,marko:98}, and although it is generally far from trivial~\cite{vkampen-lodder:84}, our results will permit taking the limit from soft to rigid constraints.

For an extensible but very stiff backbone with only small stretching deformations, the Hamiltonian reads~\cite{marko:98}
\begin{equation}\label{eq:hamiltonian}
\mathcal{H} = \frac{\kb T}{2}\int_0^L\!\!\td s\left[\lp\bvec r''^2+\ks u^2\right],
\end{equation}
where $L$ is the unstretched contour length, $\lp$ is the persistence length, $\ks$ is the stretching elastic constant, and $u=|\bvec r'|-1$ is the elongational strain. Our theory relies on the weakly-bending limit $\bvec r(s) = (s-\rpar, \rperp)^T$ of small transverse and longitudinal contour deviations $\rperp$ and $\rpar$ from a straight line, which gives $u\approx -\rpar'+\frac{1}{2}\rperp'^2$ to leading order. Observing that the polymer's longitudinal extension in the limit of large prestress $\fpre\gg \kb T/\lp$ is given by~\cite{marko:98}
\begin{equation}\label{eq:force-extension}
\frac{R_\parallel}{L} = \int_0^L\!\frac{\td s}{L} (1-\rpar') = 1+\ave{u}-\ave{\tfrac{1}{2}\rperp'^2},
\end{equation}
we can quantify the simultaneous limits of an only slightly extensible backbone and a weakly-bending contour by requiring that the contributions of longitudinal strain $\ave{u}=\fpre/(\kb T\ks)
\equiv\epsx \ll 1$ and of thermal stored length $\ave{\tfrac{1}{2}\rperp'^2}=[\kb T\lp/(4\fpre)]^{1/2}\equiv\epsth\ll 1$, respectively, are both small. Although these contributions are independent, we will also assume that the force-extension Eq.~\eqref{eq:force-extension} is dominated by contour straightening instead of backbone stretching, i.e., that $\ave{u}\ll \ave{\tfrac{1}{2}\rperp'^2}$, which is easily fulfilled as long as $\fpre\ll \fx$, where $\fx=\kb T \ks^{2/3}/\lp^{1/3}$ is the corresponding crossover force scale~\cite{wang-etal:97,kierfeld-etal:04}. This assumption is reasonable in most experimental circumstances, considering that $\fx\approx 75\pN$~\cite{kojima-etal:94} for actin, and $\fx\approx 50\pN$ for DNA~\cite{marko:98}, which in fact is close to the overstretching transition. For bead-spring simulations, however, this condition is much harder to obey~\cite{nam-lee:07},
because very small time steps $\Delta t\lesssim \ks^{-1}$ are required. Also, in special situations like the relaxation from a low-temperature initial condition~\cite{obermayer-etal:09} short-time transients are quite pronounced, and we will show below that even a prestress of only about $0.01\fx$ gives rise to observable effects.

To quantitatively assess the influence of backbone stretching on the dynamics, we proceed with a discussion of the equations of motion $\bvec\zeta\pd_t \bvec r = -\delta\mathcal{H}/\delta\bvec{r} + \bvec\xi$, with the stochastic noise $\bvec \xi$ and the free-draining friction matrix $\bvec\zeta=\zeta_\perp[\bvec r'\bvec r'+\hat\zeta(1-\bvec r'\bvec r')]$, where $\hat\zeta\approx 1/2$ accounts for the anisotropy between transverse and longitudinal friction. To leading order in $\epsth$ and $\epsx$ and in the absence of external forces, we obtain
\begin{subequations}\label{eq:eom}
\begin{gather}
\label{eq:eom-T}
\pd_t\rperp = -\rperp''''+\ks(u \rperp')'/\lp  + \bvec \xi_\perp, \\
\label{eq:eom-L}
\hat\zeta \pd_t \rpar + (1-\hat\zeta)\rperp'\pd_t\rperp = -\rpar''''-\ks u'/\lp + \xi_\parallel.
\end{gather}
\end{subequations}
Here, we have introduced units such that $\kb T\equiv\lp^{-1}$ and $\zeta_\perp=1$, which makes time a length$^{4}$ and force a length$^{-2}$. In the following, we are interested in the prototypical rheological experiment~\cite{fernandez-pullarkat-ott:06,gardel-etal:06,kasza-etal:09} where at time $t=0$ a small time-dependent force $\delta f(t)$ is superimposed on a static prestress $\fpre$, which contributes a term $[\fpre+\Theta(t)\delta f(t)][\delta(L-s)-\delta(s)]$ on the right hand side of Eq.~\eqref{eq:eom-L}. In the stationary state at times $t < 0$, this gives $\ks u/\lp=\fpre$, and the combination $\ks u/\lp$  plays the role of a tension in Eq.~\eqref{eq:eom-T} also at later times.

If we assume constant $u=\lp\fpre/\ks$, we find that the transverse part Eq.~\eqref{eq:eom-T} is correlated on length scales $\lperp(t)$ with $\lperp\sim t^{1/4}$ if $t\ll\fpre^{-2}$ and $\lperp(t)\sim (\fpre t)^{1/2}$ if $t\gg \fpre^{-2}$~\cite{everaers-juelicher-ajdari-maggs:99,hallatschek-frey-kroy:05}. On the other hand, disregarding the thermal contribution $\rpar$ to Eq.~\eqref{eq:eom-L} for the moment, we also find that the diffusive dynamics of the elongational strain $u$  is correlated on length scales $\lx(t)\sim (\ks t/\lp)^{1/2}$~\cite{marko:98}. Given now that $\lp\fpre/\ks=\mathcal{O}(\epsx)\ll 1$, it turns out that $\lx \gg \lperp$ except for veryy early times $t\lesssim (\lp/\ks)^2$, where higher order terms become relevant. Hence, after short initial transients the elongational strain $u$ varies slowly with arclength: stretching modes relax extremely fast, but only to a local equilibrium value, which is not only nonzero for a prestressed filament, but can even show nontrivial large-scale spatial variations for the previously mentioned nonequilibrium stretching experiments. Thus, elongational strain cannot globally equilibrate unless these tension variations have propagated through the filament. The latter are linked to the dynamics of thermal stored length and therefore significantly slowed down by longitudinal friction, and it has been shown that the associated characteristic length scale $\lpar(t)\propto \epsth^{-1/2}\lperp(t)$ is much larger than the one of transverse fluctuations~\cite{hallatschek-frey-kroy:05}.

Our goal is now to formulate an equation for the elongational strain $u$ that integrates over transverse fluctuations described through Eq.~\eqref{eq:eom-T} (on the short length scale $\lperp$), but retains both large-scale \emph{spatial} variations in the tension $\ks u/\lp$ (on the scale $\lpar$) as well as the effect of short \emph{time} transients stemming from the fast relaxation of elongational modes. To this end, we employ a  multiple scale perturbation theory both in space and time: small-scale and large-scale spatial coordinates $s$ and $\epsth^{1/2}\bar s$, respectively, account for the different \emph{spatial} correlation lengths of transverse and longitudinal contour displacements, while slow and fast time variables $t$ and $\tau=(\epsth/\epsx) t$ account for long time stored length and short time strain dynamics, respectively. Here, the condition $\epsx \ll \epsth$ (resulting from $\fpre\ll\fx$) is essential, and it entails that a $\tau$-derivative of $u$ is of order $\epsth$: transients from strain relaxation  can become comparable to thermal stored length. We briefly sketch the analysis, which in its technical details is quite analogous to Ref.~\cite{hallatschek-frey-kroy:07}. Taking an $s$-derivative of Eq.~\eqref{eq:eom-L} and eliminating $\rpar'=-u+\frac{1}{2}\rperp'^2$, we find to zeroth order in $\epsth$: $-\hat\zeta\pd_t u = \pd_s^4 u-\ks\pd_s^2 u/\lp$. Since $\ks u/\lp=\mathcal{O}(\fpre)$ while everything else is $\mathcal{O}(\epsx)$, we find $\pd_s^2 u=0$, and therefore also $\pd_s^4 u=\pd_t u=0$. This also implies that Eq.~\eqref{eq:eom-T} becomes the linear equation $\pd_t \rperp=-\pd_s^4 \rperp + \ks u\pd_s^2 \rperp/\lp + \bvec \xi_\perp$. To first order in $\epsth$, Eq.~\eqref{eq:eom-L} now gives $\hat\zeta\frac{1}{2}(\pd_s\rperp)^2 -\hat\zeta\epsth\pd_\tau u/\epsx = -\ks\pd_{\bar s}^2 u/\lp + H(s,\bar s)$, where $H(s,\bar s)$ summarizes $s$-derivatives of terms nonlinear in $\rperp$. These vanish upon coarse-graining, i.e., when averaging this equation over small-scale variations on the scale $\lperp$~\cite{hallatschek-frey-kroy:07}. Denoting in such a manner spatially averaged quantities with an overbar and replacing $\tau\to t$, we obtain the relation
\begin{equation}\label{eq:tension-dynamics}
\ks\pd_{\bar s}^2 \bar u/\lp = -\hat\zeta\pd_t(\bar \varrho-\bar u),
\end{equation}
where $\varrho=\frac{1}{2}\rperp'^2$ is the local density of contour length stored in thermal undulations. This relation, which is our main result, formalizes in an intuitive way the opposing effects thermal stored length density $\bar\varrho$ and elongational strain $\bar u$ have on the backbone tension $\ks\bar u/\lp$: if prestress is increased, stored length is destroyed and elongational strain created, and both lead temporarily to spatial tension inhomogeneities (curvature), and vice versa for decreasing prestress. Further, taking the ``inextensible'' limit $\epsx\to 0$ while holding the tension $\ks\bar u/\lp$ fixed leads to the inextensible analog derived in Ref.~\cite{hallatschek-frey-kroy:05}, and the limit $\epsth\to 0$ of a one-dimensional Rouse chain, although our assumptions cease to hold, nevertheless gives a simple diffusion equation for $\bar u$.

In order to solve Eq.~\eqref{eq:tension-dynamics}, we observe that its boundary conditions are prescribed through the externally applied prestress: $\bar u(0,t)=\bar u(L,t)=\lp[\fpre+\Theta(t)\delta f(t)]/\ks$. Further, $\bar\varrho$ has to be computed from Eq.~\eqref{eq:eom-T}, which to lowest order depends only parametrically on $\bar u(\bar s,t)$. It is thus effectively linear in $\rperp$ and can be solved in Fourier space by means of the response function~\cite{hallatschek-frey-kroy:07}
$\chi_\perp(q;t,t') = \e^{- q^2 [q^2(t-t') + \ks(\bar U(t)-\bar U(t'))/\lp]}$, where $\bar U(t)=\int_0^t\!\td t'\bar u(t')$ is the time-integrated strain. Because spatially averaging $\bar\varrho$ over many effectively uncorrelated segments of length $\lperp$ produces an ensemble average, the stored length density is given by~\cite{hallatschek-frey-kroy:07}
\begin{equation}\label{eq:varrho}
\bar\varrho = \int_0^\infty\!\frac{\td q}{\pi\lp}\left[\frac{\chi_\perp^2(q;t,0)}{q^2+\fpre} + 2 q^2\int_0^t\!\td t' \chi_\perp^2(q;t,t')\right].
\end{equation}
Solutions to Eq.~\eqref{eq:tension-dynamics} can now be obtained similarly to Refs.~\cite{hallatschek-frey-kroy:07,obermayer-etal:07,obermayer-etal:09}.

In the remainder of this paper, we focus on a small oscillatory stress $\delta f(t)=\delta f\sin\omega t$ superimposed on a large prestress $\fpre$. This situation has been analyzed in Ref.~\cite{hiraiwa-ohta:08} for an inextensible filament, and our calculation proceeds along these lines. While we cannot actually take the linear response limit $\delta f\to 0$ because then the assumptions underlying the multiple scale perturbation theory become invalid~\cite{hallatschek-frey-kroy:05}, we can still for small enough $\delta f\ll\fpre$ linearize Eq.~\eqref{eq:varrho} by writing $\bar U(\bar s,t)=\lp\fpre t/\ks+\Delta \bar U (\bar s, t)$ and find from Eq.~\eqref{eq:tension-dynamics} a simple equation for the Laplace transform $\Delta \tilde U(\bar s,z)$~\cite{hallatschek-frey-kroy:07,hiraiwa-ohta:08}:
\begin{equation}\label{eq:tension-dynamics-laplace}
\pd_{\bar s}^2 \Delta \tilde U (\bar s,z) = M(z) \Delta \tilde U(\bar s,z). 
\end{equation}
The kernel $M(z)=\hat\zeta\fpre^{1/2}\hat M(z\fpre^{-2})/\lp$ is defined as
\begin{equation}\label{eq:laplace-kernel}
\hat M(\hat z)=\int_0^\infty\!\frac{\td k}{\pi}\left[\frac{2 k^2}{k^2+1} - \frac{4 k^4}{2 k^2(k^2+1)+\hat z}\right]+\phix^{3/2} \hat z,
\end{equation}
and $\phix=\fpre/\fx\ll 1$ is the ratio of prestress to the critical force $\fx=\ks^{2/3}/\lp^{4/3}$.  In constrast to the inextensible case treated in Ref.~\cite{hiraiwa-ohta:08}, where $\hat M(\hat z)\propto \hat z^{1/4}$ for all $\hat z \gg 1$, we obtain now an additional high-frequency regime $\hat M(\hat z)\sim \phix^{3/2}\hat z$ for $\hat z\gg \phix^{-2}$. Considering that $M^{-1/2}(z)$ is the analog in Laplace space to the characteristic length scale $\lpar(t)$ for the large-scale spatial tension variations, we find that $\lpar(t)\simeq (\ks t/\lp)^{1/2}$~\cite{marko:98} shows a diffusive scaling in the corresponding extensibility-dominated short-time regime $t\ll\tx$,  only then crosses over to the well-known growth law $\lpar(t)\simeq \lp^{1/2} t^{1/8}$~\cite{everaers-juelicher-ajdari-maggs:99}, before finally arriving at $\lpar(t)\simeq \lp^{1/2}\fpre^{3/4} t^{1/2}$~\cite{brochard-buguin-degennes:99,hallatschek-frey-kroy:05} for $t\gg\fpre^{-2}$ (i.e., $\hat z\ll 1$). We emphasize that the scaling of the crossover time $\tx=\lp^{8/3}/\ks^{4/3}=\fx^{-2}$ cannot simply be inferred from dimensional analysis, because any combination of the lengths $\lp$ and $\ks^{-1}$ could be used.

The observable of main interest is the change $\delta R_\parallel(t)$ in projected length $R_\parallel$, which is through the force-extension relation Eq.~\eqref{eq:force-extension} related to the integrated change of $\bar u-\bar \varrho$ and thus through Eq.~\eqref{eq:tension-dynamics} to the features of the tension $\ks \bar u/\lp$:
\begin{equation}\label{eq:observable}
\delta R_\parallel(t) = \int_0^t\!\!\td t'\!\int_0^L\!\!\td s \,(\bar u-\bar \varrho)=\frac{\ks}{\hat\zeta\lp} [\pd_{\bar s}\bar U(L,t)-\pd_{\bar s}\bar U(0,t)].
\end{equation}
In Laplace space, we thus obtain from a straightforward solution of Eq.~\eqref{eq:tension-dynamics-laplace} $\delta \tilde R(z)=\frac{2}{\hat\zeta z}M^{1/2}(z)\times\tanh[\frac{L}{2}M^{1/2}(z)]\delta \tilde f(z)$, which can be backtransformed in the stationary limit $t\to\infty$~\cite{hiraiwa-ohta:08}:
\begin{equation}\label{eq:result-deltar}
\delta R_\parallel(t) = \frac{\delta f}{\hat \zeta L \fpre^2 }\left[\hat J'(\omega \fpre^{-2})\sin(\omega t)-\hat J''(\omega\fpre^{-2})\cos(\omega t)\right],
\end{equation}
with the dimensionless compliances
\begin{equation}\label{eq:compliances}
\begin{split}
\hat J'(\hat \omega) &= \frac{2}{\hat \omega} \mathrm{Im} \left[\sqrt{\phic^{1/2}\hat M(\imi \hat\omega)} \tanh\sqrt{\tfrac{1}{4}\phic^{1/2}\hat M(\imi\hat\omega)}\right], \\
\hat J''(\hat \omega) &= \frac{2}{\hat \omega} \mathrm{Re} \left[\sqrt{\phic^{1/2}\hat M(\imi \hat\omega)} \tanh\sqrt{\tfrac{1}{4}\phic^{1/2}\hat M(\imi\hat\omega)}\right]. 
\end{split}
\end{equation}
Here, $\phic=\fpre/\fc$ gives the ratio of $\fpre$ to the longitudinal critical force $\fc=\lp^2/(\hat\zeta^2 L^4)$ and can be used to distinguish ``long'' ($\phic\gg 1$) and ``short'' ($\phic\ll 1$) filaments, respectively~\cite{obermayer-etal:07}.

\begin{figure}
\includegraphics[angle=270,width=.48\textwidth]{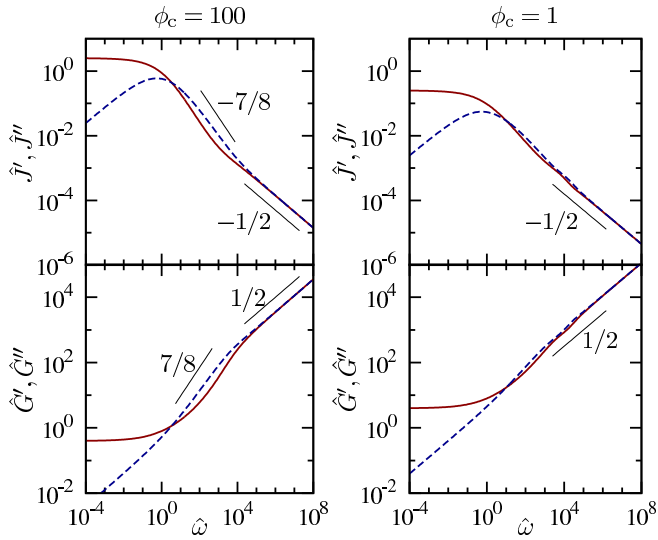}
\caption{\label{fig:compliances}Plot of the moduli $\hat J',\hat J''$ (top) and $\hat G',\hat G''$ (bottom) from Eq.~\eqref{eq:compliances} for $\phix=\fpre/\fx=10^{-2}$ and for $\phic=100$ (left) and $\phic=1$ (right), where $\phic=\fpre/\fc$ with $\fc=\lp^2/(\hat\zeta^2 L^4)$ and $\fx=\ks^{2/3}/\lp^{4/3}$. Solid lines: $\hat J'$ ($\hat G'$), dashed lines: $\hat J''$ ($\hat G''$). Short lines indicate high-frequency scaling laws of Table~\ref{tb:compliances}.}
\end{figure}

Fig.~\ref{fig:compliances} depicts numerical solutions of Eq.~\eqref{eq:compliances} for a fixed value of $\phix=10^{-2}$ and different values of $\phic$, as well as results for the corresponding viscoelastic elastic modulus $(\hat G'+\imi \hat G'')=(\hat J'+\imi \hat J'')^{-1}$. It is straightforward to check that Eq.~\eqref{eq:compliances} obeys scaling laws in different intermediate asymptotic regimes, which are summarized in Table~\ref{tb:compliances}. Most of these results have been obtained previously (cf.~Refs.~\cite{morse:98c,gittes-mackintosh:98,granek:97,caspi-etal:98,everaers-juelicher-ajdari-maggs:99,hiraiwa-ohta:08}), and we will therefore not comment on them in detail. We merely emphasize that for low frequencies $\hat\omega \lesssim 1$ the presence of a prestress leads to the well-known nonlinear response regime with its characteristic $\frac{1}{2}$-exponents~\cite{caspi-etal:98,mizuno-etal:07}. The intermediate regime $\phic^{1/2}\ll\hat\omega\ll 1$ with $\hat J'\sim\hat J''\sim \hat\omega^{-1/2}$  corresponds to the previously discussed regime of nonlinear tension propagation~\cite{brochard-buguin-degennes:99,hallatschek-frey-kroy:05} and can only be observed for large $\phic\gg 1$. Also, for small prestress ($\phic\ll 1$), there is an intermediate regime with $\hat J'\sim\hat J''\sim \hat \omega^{-3/4}$ for $\phic^{-2}\gg\hat\omega\gg 1$ equivalent to the force-free case~\cite{morse:98c,gittes-mackintosh:98}.

\begin{table}
\begin{tabular*}{.48\textwidth}{@{\extracolsep{\fill}}ccc}
(a)  & $\hat J'(\hat\omega)$ & $\hat J''(\hat\omega)$ \\
\hline\hline
$\hat\omega\gg\phix^{-2}$ & \multicolumn{2}{c}{$\phic^{1/4}\phix^{3/4}\hat\omega^{-1/2}$} \\
$\phix^{-2}\gg\hat\omega\gg 1$ & \multicolumn{2}{c}{$\phic^{1/4}\hat\omega^{-7/8}$} \\
$1\gg\hat\omega\gg \phic^{-1/2}$ & \multicolumn{2}{c}{$\phic^{1/4}\hat\omega^{-1/2}$} \\
$\phic^{-1/2}\gg \hat\omega$ & $\phic^{1/2}$ & $\phic^{1/2}\hat\omega^{1/2}$
\end{tabular*}

\begin{tabular*}{.48\textwidth}{@{\extracolsep{\fill}}ccc}
(b)  & $\hat J'(\hat\omega)$ & $\hat J''(\hat\omega)$ \\
\hline\hline
$\hat\omega\gg\phix^{-2}$ & \multicolumn{2}{c}{$\phic^{1/4}\phix^{3/4}\hat\omega^{-1/2}$} \\
$\phix^{-2}\gg\hat\omega\gg \phic^{-2}$ & \multicolumn{2}{c}{$\phic^{1/4}\hat\omega^{-7/8}$} \\
$\phic^{-2}\gg\hat\omega\gg 1$ & \multicolumn{2}{c}{$\phic^{1/2}\hat\omega^{-3/4}$} \\
$1\gg \hat\omega$ & $\phic^{1/2}$ & $\phic^{1/2}\hat\omega^{1/2}$
\end{tabular*}
\caption{\label{tb:compliances}Asymptotic scaling results~\cite{morse:98c,gittes-mackintosh:98,granek:97,caspi-etal:98,everaers-juelicher-ajdari-maggs:99,hiraiwa-ohta:08} for the compliances $\hat J'(\hat\omega)$ and $\hat J''(\hat\omega)$ from Eq.~\eqref{eq:compliances} with $\hat\omega=\omega\fpre^{-2}$ for $\phic\gg 1$ (a) and $\phic\ll 1$ (b), respectively, where $\phix=\fpre/\fx$ and $\phic=\fpre/\fc$ with $\fx=\ks^{2/3}/\lp^{4/3}$ and $\fc=\lp^2/(\hat\zeta^2 L^4)$.}
\end{table}

The predominant effect of an extensible backbone is to produce a Rouse-like scaling in the new high-frequency regime $\omega \gg \fx^{2}$. Here, we find 
$\hat J'(\hat \omega)\sim\hat J''(\hat \omega) \sim 2^{1/2} \phic^{1/4}\phix^{3/4}\hat\omega^{-1/2}$, corresponding to $\hat G'\sim \hat G'' \sim 2^{-3/2} \phic^{-1/4}\phix^{-3/4}\hat\omega^{1/2}$. From Fig.~\ref{fig:compliances} we conclude that a moderate prestress $\fpre$ of merely 1\% of the critical force $\fx$ suffices to significantly shorten the $7/8$-regime and to produce a distinct Rouse-like signature, especially if $\phic$ is not too large.

In summary, we have presented a systematic theory of tension dynamics for extensible wormlike chains including the opposing effects of thermal stored length and elongational strain relaxation at short times. These produce a Rouse-like scaling in the high-frequency viscoelastic response, and are expected to be especially relevant for the proper design of bead-spring simulations.

\acknowledgments
We gratefully acknowledge helpful discussions with Klaus Kroy and Oskar Hallatschek, and financial support from Deutsche Forschungsgemeinschaft through SFB 486 and by Nanosystems Initiative Munich.

\end{document}